\documentclass[journal]{IEEEtran} 
\usepackage[T1]{fontenc}
\usepackage[utf8]{inputenc}

\usepackage{amsmath, amssymb, amsfonts}
\usepackage{amsthm}

\usepackage{graphicx}
\usepackage{subcaption}

\usepackage{array}
\usepackage{tabularx}
\usepackage{booktabs}
\usepackage{multirow}
\usepackage{makecell}
\usepackage{cellspace}

\usepackage[table, svgnames, dvipsnames]{xcolor}

\usepackage{algorithm}
\usepackage{algorithmic}

\usepackage[shortlabels]{enumitem}

\usepackage{url}
\usepackage{hyperref}
\usepackage{float}
\usepackage{cite}
\usepackage{comment}
\usepackage{soul}
\usepackage{xspace}
\usepackage{slashed}
\usepackage{orcidlink}

\definecolor{maroon}{cmyk}{0,0.87,0.68,0.32}
\sethlcolor{gray}

\begin{document}

\title{Statistical Analysis of the Reliability of Data Collected with Wireless Electrocardiograms Outside Clinical Settings}
\author{Yalemzerf Getnet~\orcidlink{0009-0008-5975-2086}, Waltenegus Dargie~\orcidlink{0000-0002-7911-8081}, \IEEEmembership{Senior Member, IEEE}  
    \thanks{Manuscript submitted on 13 November 2025.}
    \thanks{Y. Getnet is with the Department of Electrical and Computer Engineering at Addis Ababa University, Ethiopia (e-mail:  yalemzerf.getnet@aau.edu.et)}
    \thanks{ W. Dargie is with the Faculty of Computer Science, Technische Universit{\"a}t Dresden, 01062 Dresden, Germany (waltenegus.dargie@tu-dresden.de)}
}
\maketitle
\begin{abstract}
Cost-effective wireless electrocardiograms (ECGs) enable long-term and scalable monitoring of cardiac patients in their home and work environments. Because they offer greater freedom of movement, they are also suitable for investigating the relationship between cardiac workload and underlying physical exertion. However, this requires that the quality of the generated data meets the standards of clinical devices. The aim of this study is to examine this closely. We therefore analyze data from 54 healthy subjects who performed five physical activities using wireless ECGs outside of clinical settings and without medical supervision. The results are compared with clinically collected data from standard 12-lead ECGs (2493 subjects) and Holter ECGs (29 subjects), with particular attention to the RR interval time series (tachogram) and heart rate variability (HRV). Our study shows significant statistical agreement between the different datasets. We calculated the 95\% confidence intervals for the mean RR interval and HRV assuming that (1) the statistics of the 12-lead ECGs could serve as reliable reference, and (2) the statistics of the 12-lead ECGs cannot be taken as reliable reference. The p-values for both conditions (for the RR interval: 0.23 and 0.26 respectively; for HRV: 0.10 and 0.11 respectively) suggest that there is insufficient evidence to reject the hypothesis that significant statistical agreement exists between the different datasets.
\end{abstract}

\section{Introduction}
\label{sec: introduction} 

Research into wireless (wearable) electrocardiograms (ECGs) and their use outside of clinical environments for monitoring various heart diseases is ongoing \cite{masihi2021development, chen2022contactless}. Combined with state-of-the-art machine learning models, these devices promise cost-effective, long-term, and scalable monitoring \cite{farooq2021wearable, farrokhi2024human, strodthoff2020deep}. This capability not only reduces the complexity and cost of cardiac treatment but also opens up new possibilities for preventive healthcare \cite{gomez2021platform, hussain2021big}. However, there remain some challenges for cardiologists to rely on such devices to make clinically relevant and binding decisions. Since the ECG recording is a legally valid document \cite{breen2022ecg, houghton2025making}, the data generated by the devices must be of the same quality as clinically certified counterparts \cite{bayoumy2021smart}. Secondly, whereas the medical devices used in clinical settings are operated by medically trained personnel and the patients' physical conditions are generally known and monitored, the same preconditions cannot be guaranteed when wireless electrocardiograms are used by laypersons outside of clinical settings.

In this paper we investigate the quality of data generated by different types of electrocardiograms in different settings and compare their essential statistical features, paying particular attention to the RR-interval and heart rate variability (HRV), as these features are crucial for the study of various cardiac conditions as well as for testing cardiac fitness \cite{fujiwara2018heart, dogan2025continuous}. To make the comparison objective, we focus on healthy subjects. Furthermore, the paper examines the correlation between physical activities (exertions) and cardiac responses, from which the dynamics between sympathetic and parasympathetic activity can be inferred \cite{dogan2025continuous}. The relationship has so far been insufficiently addressed in the literature, as actions related to the sympathetic nervous system are often induced by stress or medication to trigger the hypothalamus to send signals to the adrenal glands to release epinephrine (adrenaline) and other hormones into the bloodstream \cite{elghozi2007sympathetic, rohleder2004psychosocial}. Wireless electrocardiograms, on the other hand, allow for greater freedom of movement; thus enabling the observation of significant HRV through variations in physical activity levels and the objective investigation of their correlation \cite{fleshner2005physical}.


Our study is based on both proprietary and publicly available data. The first dataset consists of ECG data collected from 54 participants using a low-power, 5-lead wireless ECG while the participants performed five different physical activities. The second dataset consists of  publicly available ECG data from 2,493 participants recorded with standard 12-lead ECGs in controlled clinical environments. The third dataset consists of data collected from 29 participants using a medically certified wireless ambulatory Holter ECG.

The specific research questions we pose and aim to answer are stated as follows:
\begin{itemize}
\item How significant is the statistical overlap between the data collected in non-clinical settings and the data collected in clinical settings or under clinical supervision?
\item In the study of cardiac fitness, understanding the relationship between sympathetic and parasympathetic activity (the so-called sympathovagal balance \cite{strigo2016interoception, di2016does}) is important. How well can this balance be observed when a wireless ECG is used outside of a clinical setting and subjects carry out different types of physical activities?
\end{itemize}

Our study suggests that both questions can be answered in the affirmative. The interval estimates for the means and variances of the RR interval time series and HRV yield p-values that  exceed the significance level.

The remainder of the paper is organized as follows: In Section \ref{sec:related}, we review related work. In Section \ref{sec:methodology}, we describe the data acquisition and discuss preprocessing. In Sections~\ref{sec:stats}, we discuss the statistical significance of the different datasets and compare essential statistical and ECG features. In Section~\ref{sec:estimation}, we compute the confidence interval for the different ECG parameters we discuss in Section~\ref{sec:stats} setting different preconditions. In Section~\ref{sec:hrv}, we analyze the relationship between physical exertion and cardiac workload. Finally, in Section \ref{sec:conclusion}, we provide concluding remarks and outline future work.   

\section{Related Work}
\label{sec:related}
Various studies have explored the relevance of wearable ECG devices to monitor cardiac activity outside of clinical environments \cite{bayoumy2021smart,fujiwara2018heart}.  Most of the studies conducted demonstrate that wearable ECG devices are feasible to be used in continuous cardiac monitoring \cite {fuller2020reliability,  zang2025novel}. 
The study conducted in \cite{rafols2018evaluation} analyzed the reliability of a Shimmer (v. 3) wearable device to extract cardio-respiratory parameters. Twenty  healthy subject (13 males and 7 females,  aged 22.70 ± 1.53 ) were involved in the data collection process. The heart rates of the subjects were estimated from cardiorespiratory parameters and the result were compared with those estimated with a standard MP150. The wireless ECG achieved an overall HR detection rate of 97.32\% which was nearly the same as the result obtained with the reference device (97.56\%).

The authors in \cite{lazaro2020wearable} developed a  heart rate measuring electronic armband.  ECG data were recorded from 16 healthy subjects (age: 27.56 ± 8.82 ) using  the device and a Holter ECG. Four classical HRV parameters (SDNN, RMSSD, and powers at low and high frequency bands) were computed. During non-bedtime periods, 75.25\%  of the heart rate measurements of the armband is within $\pm $10\% of the Holter mean heart rate. During bedtime periods, the accuracy improved from 75.25\% to 98.49\%.

The study in \cite {bent2020investigating} evaluated the accuracy of consumer-grade and research-grade optical wearable devices in estimating Heart rate and PPG. Data were collected from 53 subjects (32 females, 21 males; ages 18-54),  using four consumer-grade optical devices (FITBIT, GARMIN, BIOVOTION and EMPATICA) and two research-grade optical devices (Miband and the Apple Watch 4). The wearable devices HR readings were compared to HR values calculated from simultaneously measured electrocardiogram (ECG). At resting condition, the MAE of consumer-grade wearable devices was  7.2 ± 5.4 bpm and the MAE of research-grade wearable devices was 13.9 ± 7.8 bpm. During physical activity, the MAE  of consumer wearable devices was 10.2 ± 7.5 bpm and the MAE of research-grade wearable devices was 15.9 ± 8.1 bpm.

In  \cite{natarajan2020heart} the authors analyzed HRV to understand  how it is affected by different factors, including age, sex, and physical activity. Inter-beat interval collected from 8.2 million Fitbit wrist-worn device users were used to analysis HRV. Five min windows were considered for the time and frequency domain metrics and 60 min measurements were used for graphical domain metrics. The result obtained show that the SDRR, low-frequency power, and Poincare S2 (to be discussed in Section~\ref{sec:hrv}) vary significantly  with sex; however no appreciable difference was seen with RMSSD, high-frequency power, and Poincare S1 across sex.

The study conducted in \cite{betti2017evaluation} evaluated the reliability of the BioHarness 3, Shimmer Sensor and MindWave Mobile EEG headset, based on ECG, EDA, and EEG signals. Data were collected from 15 subjects (7 females and 8 males,  mean age 40.8 ± 9.5 years) while they were under stress.  The SVM classifier was used for identification of stress, which it carried out with an accuracy of 86.0\% accuracy.
 
The study in \cite{8638779} estimated driver fatigue using  frequency-domain and non-linear metrics derived from contactless ECG (cECG).  A non-contact capacitive coupled electrocardiography was developed and used to collect data from 20 male volunteers (age = 23±3.1 yrs).  Correlation between frequency indices and non-linear indices of HRV was analyzed, and the result showed that the correlation between LF and non-linear indices of cECG signals was not significant. However, the correlation  was found to be significant for some frequency-domain and non-linear metrics such as HF with SD1,  HF with SD1/SD2,  LF/HF with  SD1 and SD1/SD2 at the significance level of p $< $0.05. 

The study in \cite{li2025motion} proposed a motion-unrestricted dynamic electrocardiogram (MU-DCG), which employs skin-conformal, imperceptible electronics for long-term accurate 12-lead ECG monitoring. The dynamic 12-lead ECG signal were recorded from 3 healthy subjects (all male, aged 23, 24, and 26) performing various physical activities, including sitting , walking, jogging, ascending and descending stairs.  The proposed system achieved a higher signal-to-noise ratios ($>$ 23 dB ) compared with conventional Holter device ($<$ 17dB) during walking, jogging, ascending and descending stairs.
\begin{table*}[ht]
\centering
\scriptsize
\caption{Comparison of studies on the reliability of wireless ECG signals}
\label{tab:related_work_comparison1}
\begin{tabular}{|l|l|l|l|l|l|l|l|l|}
\hline
\textbf{\parbox{0.5cm}{ Work}} &
\textbf{\parbox{0.8cm}{Dataset}} &
\textbf{\parbox{1cm}{Condition}} &
\textbf{\parbox{0.8cm}{No. of\\subjects}} &
\textbf{\parbox{1cm}{Device}} &
\textbf{\parbox{1cm}{ Model, Metrics}} &
\textbf{\parbox{1.5cm}{ Performance}} &
\textbf{\parbox{1.5cm}{ Uniqueness / Contribution}} &
\textbf{\parbox{2.8cm}{ Remarks / Limitations}} \\
\hline

\cite{rafols2018evaluation} & Proprietary & controlled & 20 & \parbox{1cm}{ Shimmer (v.3)\\MP150} & \parbox{0.8cm}{ Pan-Tompkins \\algorithm} & \parbox{2.5cm}{Sensitivity: 97.56\% (MP150)\\97.32\% (Shimmer)} &Comparison  & \parbox{2.8cm}{ HR estimated from EMG; comparison with directly measured clinical ECG} \\
\hline
\cite{lazaro2020wearable} & Proprietary & uncontrolled & 16 & armband& \parbox{1.3cm}{SDNN \&\\ RMSSD }& \parbox{2.5cm}{accuracy:  75.25\% \\(non-bed-time); 98.49\% (bed bed-time) } & \parbox{1.5cm}{design a wearable armband }& \parbox{2.8cm}{ no information about the physical activities }\\
\hline
\cite{bent2020investigating} & Proprietary & uncontrolled & 53 &\parbox{1cm} {optical HR sensors} & MAE & \parbox{2.5cm}{ At rest MAE: 7.2 $\pm$ 5.4 bpm (consumer-grade) 13.9 $\pm$ 7.8 bpm (research-grade)} &\parbox{2cm}{Compared 6  wearable devices } &
\parbox{2.8cm} {Study did not \\specify the device used \\to record reference ECG}\\
\hline
\cite{natarajan2020heart} & \parbox{0.8cm} {Fitbit\\ Research database} & uncontrolled & 8203261 & \parbox{1cm}{Fitbit wristband} & Correlation & \parbox{2.5cm}{RMSSD \& S1: 0.916\\SDRR \& S2: 0.617} & Large dataset
& \parbox{2.8cm} { Physical activity estimated using average daily steps} \\
\hline

\cite{betti2017evaluation} & Proprietary & controlled & 15 & \parbox{1cm}{wearable sensor }& SVM & Accuracy: 86\% &\parbox{2cm}{3 wearable sensors for stress measurement} & 
\parbox{2.8cm}{ Validation with biological markers; not compared with medical-grade stress-monitoring devices}  \\
\hline

\cite{8638779} & Proprietary & controlled & 20 & \parbox{1cm}{non-contact ECG} & Correlation & \parbox{2.5cm}{Correlation between ECG and cECG signals: 81.4 $\pm$ 2.98\% }& \parbox{2cm}{ Compare conventional ECG and non-contact ECG} & \parbox{2.8cm}{Sensor sensitive to noise methodology not discussed} \\
\hline  

\cite{li2025motion} & Proprietary & uncontrolled & 3 & \parbox{1cm}{MU-DCG system }& SNR & \parbox{2.5cm}{Signal-to-noise ratio: >23 dB }& MU-DCG system  & Limited dataset \\
\hline

\rowcolor{LightGray}
\parbox{0.5cm}{This\\ work} & \parbox{0.8cm}{Proprietary,\\PTBXL\\RR-interval time series} & \parbox{1cm}{uncontrolled\\controlled \\controlled} & \parbox{0.8cm}{54, 2493, 29} & \parbox{1cm}{Shimmer\\ 12-lead  Holter} & \parbox{1cm}{Comprehensive\\ statistical analysis} & \parbox{2cm}{ Significant overlap in RR-interval, HRV, and physical exertion vs. cardiac response statistics} & \parbox{2cm}{Comparison of three heterogeneous datasets }  & \parbox{2.8cm}{ Separation of various sources of errors is challenging}

\\
\hline

\end{tabular}
\end{table*}

\section{Methodology}
\label{sec:methodology}
\subsection{Data- Acquisition}
Three different datasets are used in this study. The first consists of data collected without the supervision of a clinician or a medically trained personnel. The Shimmer platform (version 3)\footnote{\url{https://shimmersensing.com/product/consensys-ecg-development-kits/}.} \cite{burns2010shimmer} was used for data collection. Fifty-four subjects participated in the data collection. These subjects were given the wireless ECGs along with a protocol specifying the different activities (sitting, standing,  climbing up stairs, climbing down stairs, and walking) they should carry out, along with the duration of each activity (2 minutes). Otherwise, the subjects placed and operated the devices on their own.  Each device avail five ECG channels, all of which were sampled synchronously at a rate of 512 samples per second. Thus, from each individual experiment we collected ca. 61440 samples. The measurements were taken in four separate batches. The first batch took place in 2019, with 8 healthy subjects (all males, mean age = 30 yrs, SD = 6 yrs). The second and the third batches took place in 2024. The second batch consisted of 16 subjects, 11 of which were females and 5, males. For this batch, the mean age $=$ 27 yrs and SD $=$ 13 yrs. Thirteen of the subjects were healthy; one of them had asthma, another took regular medication which could affect blood pressure; and one of them, a 27 years old female, was a chain smoker. The third batch consisted of 10 healthy subjects, five females and 5 males, all between 21 and 24 years of age. The mean age was 22 yrs and SD $=$ 1.9 yrs. The fourth batch took place in 2025 and consisted of 20 subjects. The average age in this batch is 24.2 with a SD $=$2.26 yrs. All data were collected with the permission of the TU Dresden's Ethic Committee  (under Application No. EK271072017). Full consent from all participants had  been obtained prior to the experiments. 

The second dataset--PTB-XL a large publicly available electrocardiography dataset \cite{strodthoff2020deep,wagner2020ptb}--is a public dataset hosted on the PhysioNet platform\footnote{https://www.nature.com/articles/s41597-020-0495-6.}. It was collected from 21,837 subjects, aged from 0 to 95; 18,251 with heart disease and 9,514 healthy individuals. The data were collected under clinical supervision with standard 12-lead electrocardiograms, which were sampled at  500 Hz for a duration of 10 seconds per session.  The present study takes the ECG  recordings of 2,493 healthy individuals aged between 20 and 50 years.

The third dataset\footnote{https://physionet.org/content/rr-interval-healthy-subjects/}  consisted of the RR-interval time series (tachometer) of 147 healthy subjects between the ages of 1 month and 55 years, extracted from ECG data gathered with ambulatory  Holter wireless electrocardiograms. In this study the data collected from 29 subjects aged between 10 and 55 is considered.

\subsection{Pre-processing}
A bandpass filter with cutoff frequencies 0.5 and 40 Hz was applied on the raw ECG signals to suppress low-frequency baseline drifts and high-frequency noise while preserving the mormphologies of QRS complexes. In addition, a notch filter (50/60 Hz) was used to suppress power line interferences. Since baseline references vary from individuals to individuals, we normalized the raw data with Min-Max Normalization to make sure that all the measurements have the same range, namely, from 0 to 1. 

Finally, R-peaks of self collected dataset and PTB Diagnostic ECG Database were detected using the Neurokit2 Python library, a comprehensive  toolkit for  processing  a variety of biomedical signals such as ECG, photoplethysmography, and electromyography \cite{makowski2021neurokit2}.  The NeuroKit2 algorithm  relies on the QRS  detection and delineation algorithms  proposed by Martinez et al. \cite{martinez2004wavelet}, which employ a discrete wavelet transform to localize QRS peaks and to detect the local maxima associated with the QRS peaks  \cite{sangha2022automated}.

\section{Statistics}
\label{sec:stats}
The R-peak is the most prominent feature in the ECG, the least susceptible to motion artifacts, and robust in the presence of intense physical exertion. The RR interval, which can be determined by measuring the time between successive R-peaks, is useful for assessing HRV and various heart conditions, including heart failure, coronary artery disease (CAD), diabetes mellitus, hypertension, and sudden cardiac death (SCD) \cite{dogan2025continuous}. Other heart conditions, such as atrial and ventricular tachycardia and bradycardia, can also be determined by evaluating the RR interval \cite{rahul2021improved}. Likewise, the instantaneous heart rate can be determined by taking the reciprocal of the RR-interval in units of beats per minute and reveals important insights into the so-called sympathovagal balance: increased heart rate may signify slow acting sympathetic activity, whereas decreased heart rate may signify a fast acting parasympathetic (vagal) activity \cite{mcsharry2003dynamical}. 

In the following, we will use the RR interval and HRV statistics to compare the three datasets and thus evaluate the reliability of the data acquired using the wireless electrocardiograms.

\begin{figure}[H]
\centering
\includegraphics[width=0.48\textwidth]{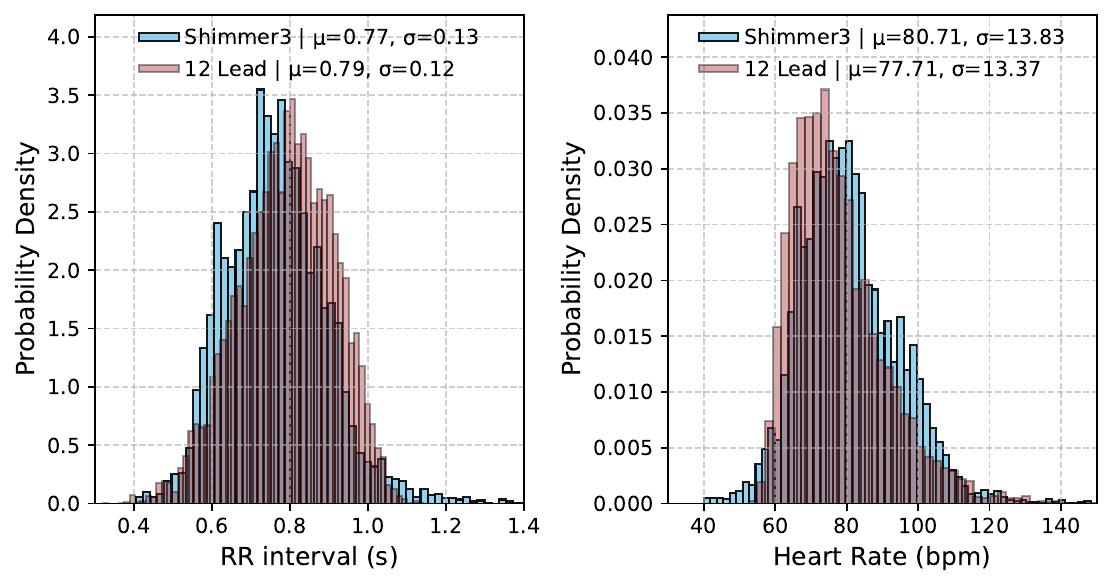}
\caption{ RR  interval and HR for  ECG signal of Shimmer3 and 12 lead}
\label{fig:rr-hr}
\end{figure}

\begin{figure}[H]
    \centering
\includegraphics[width=0.48\textwidth]{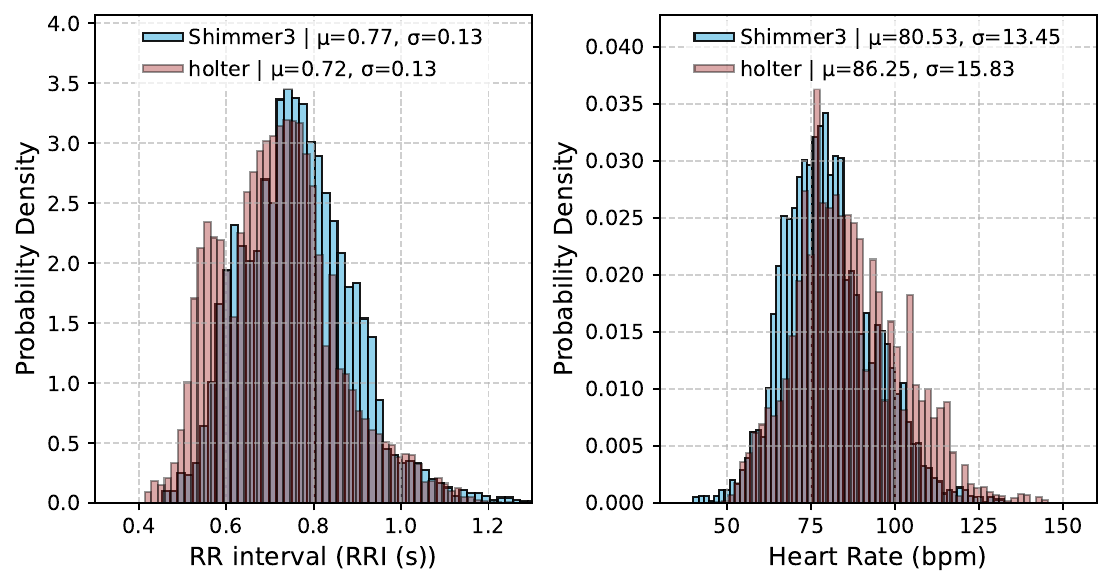}
\caption{ RR  interval and HR for  ECG signal collected using Holter ECG (2–3 leads)}
\label{fig:RRHRshimmerholter}
\end{figure}

Figs. ~\ref{fig:rr-hr} and  ~\ref{fig:RRHRshimmerholter}  compare the histograms of the RR-interval time series (RR tachogram) and HRV of the three datasets. Fig.~\ref{fig:rr-hr} compares the statistics of data collected by the Shimmer wireless ECGs when the subjects were in a relaxing state (sitting) with the data collected by the 12-lead clinical ECGs in the PTB dataset. As can be seen, there is a significant overlap both in the case of the RR-interval time series and the HRV. The RR-interval of the wireless ECG has a mean of 0.77 s and a standard deviation of 0.13 s, whereas the 12-lead ECGs statistics result in a mean of 0.79 s RR-interval with a standard deviation of 0.12 s. Even though there is a difference of ca. 0.02 s between the two means, their standard deviations suggest that with the proper calibration, the difference in the mean RR-interval becomes negligible. The same can be said of the overlap between the heart rate statistics of the wireless ECGs and the Holter ECG datasets. Here, the RR-interval from the Holter ECG has a mean of 0.72 s and a standard deviation of 0.13 s. The difference between the means of the wireless ECGs and the Holter ECGs is 0.05 s. The corresponding difference between their standard deviation is 0 s. The difference between the means of the Holter ECG and the 12-lead ECG statistics is 0.07 s. The corresponding difference between their standard deviation is 0.01. 

Likewise, the heart rate histograms suggest that the difference between the means of the wireless ECGs and the 12-lead ECGs is 3 beats per minute; the difference between the corresponding standard deviations is 0.46 beats per minute. A comparison with the statistics of the Holter ECGs suggests that the difference between the means is 5.72 beats per minute, whereas between the standard deviations is 2.38 beats per minutes. By contrast, the difference between the means of the heart rate statistics of the 12-lead ECGs and the Holter ECGs is 8.54 beats per minute, and the difference between the corresponding standard deviations is 2.46 beats per minutes. From this evaluation, it can be concluded that the statistics we established with the measurements of the wireless ECG can be taken as reliable.

\section{Parameter Estimation}
\label{sec:estimation}

Device imperfection and human physiological differences make it challenging to establish representative statistics for the ECG features we presented in Section~\ref{sec:stats}. But having representative statistics is vital for many pretraining models and data preprocessing. For example, most outlier detection algorithms rely on interquartile range  (IQR) and confidence intervals \cite{yang2021mean, blazquez2021review}. Even when machine learning algorithms are not employed, cardiologists often rely on thresholds to differentiate normal conditions from abnormal conditions as well as between different abnormal cardiac conditions \cite{Dogan10921692, monkaresi2013machine}. Probability distribution and density functions are the basis for establishing various statistical parameters (mean, variance, confidence intervals) and relationship between random variables (such as correlations, conditional means, and covariance). In order to account for the variations in statistics arising from device imperfections, an important issue to address would be how reliable are the samples we collect to determine cardiac conditions. 

In this section, we attempt to determine confidence intervals by accepting different underlying conditions. More specifically, we wish to determine how much confidence we can attach to the means and variances of the features we statistically estimated in Section~\ref{sec:stats}. Since we have no reference statistics to rely on, we compute the confidence intervals for the following conditions:
\begin{enumerate}
    \item We determine the confidence interval for the means in terms of the samples we collected using the wearable ECGs, assuming that we can take the variances of the 12-lead ECGs as reliable references.
    \item We determine the confidence interval of the means in terms of the samples without making any assumption about the variances.
      \item We determine the confidence interval for the variances in terms of the samples,  assuming that we can take the means of the 12-lead ECGs as reliable references.
    \item We determine the confidence interval for the variances in terms of the samples without making any assumption about the means.
\end{enumerate}
The sample mean and the sample variance, both of which are modeled as random variables (because every time we take samples, we may get different results), are by definition expressed as follows:

\begin{equation}
    \label{eq:p1}
    \widetilde{\mathbf{x}} = \frac{1}{n} \sum_{i = 1}^n \mathbf{x}_i
\end{equation}
The unknown random variable  $\mathbf{x}$ is the ECG feature of interest (such as HRV) and we wish to establish the confidence with which the sample random variables $\mathbf{x}_i$ represent $\mathbf{x}$. Assuming that the $\mathbf{x}_i$ are independent and identically distributed (i.i.d.) (as we are concerned with healthy people) and their mean overlaps with the mean of $\mathbf{x}$, then $\widetilde{\mathbf{x}}$ has the same mean as $\mathbf{x}$ and its variance is $\sigma^2/n$, where $\sigma^2$ is the variance of $\mathbf{x}$\footnote{In this subsection, we represent random variables with boldface letter, whereas the value which can be assigned to them are represented by regular fonts.} \cite{papoulis2002probability}. Likewise, the sample variance is given as:

\begin{equation}
    \label{eq:smean}
  \mathbf{s}^2 = \frac{1}{n-1} \sum_{i = 1}^n \left ( \mathbf{x}_i -  \widetilde{\mathbf{x}} \right )^2
\end{equation}

\begin{table}[H]
\centering
\begin{tabular}{c|c|c|l}
\hline
Parameter & Mean & Variance & Remark \\
\hline \hline
$\mathbf{x}$ & $\eta$ & $\sigma^2$ & \parbox[t]{4cm}{The unknown parameter we wish to determine} \\
\hline
$\mathbf{x}_i$ & $\eta$ & $\sigma^2/n$ & \parbox[t]{4cm}{The $n$ samples we collect using a wireless ECG} \\
\hline
$\mathbf{s}^2$ & $\sigma^2$ & 0 & \parbox[t]{4cm}{The sample variance (assuming $\eta$ is unknown)} \\
\hline
$\mathbf{v}^2$ & $\sigma^2$ & 0 & \parbox[t]{4cm}{The sample variance (assuming $\eta$ is known)} \\
\hline
\end{tabular}
\caption{Parameters of interest.}
\label{tab:parameter}
\end{table}
    
The histograms we displayed in Section~\ref{sec:stats} suggest that for large $n$ (the number of subjects), the RR-interval and HRV, expressed as two random variables, have normal distributions. This is in accordance with the Central Limit Theorem \cite{papoulis2002probability}. Therefore, in the subsequent subsections we assume that the unknown random variable $\mathbf{x}$ is normally distributed, though its mean and variance may not be known. Therefore, the confidence interval can be expressed in terms of the probability distribution and density functions, respectively: $F(\theta_2) - F(\theta_1) = \int_{\theta_1}^{\theta_2} f(x) \; dx$. Because of the normality assumption, the integration yields:
\begin{equation}
    \label{eq:p3}
   \gamma =  \int_{\theta_1}^{\theta_2} \frac{1}{\sqrt{2 \pi} \sigma} e^{-\frac{(x-\eta)^2}{2\sigma^2}} dx
\end{equation}
The integration can be simplified by normalizing the unknown random variable: $z = (x - \eta)/\sigma$, thus becoming the integration of a standardized normal distribution (with zero mean and standard deviation of one) and the interval estimation becomes:
\begin{equation}
    \label{eq:p4}
   \gamma  = \int_{(\theta_1 - \eta)/\sigma}^{(\theta_2 - \eta)/\sigma} \frac{1}{\sqrt{2 \pi}} e^{-z^2/2} dz  
\end{equation}
Thus, given a standardized normal random variable $\mathbf{z}$ and a confidence coefficient, $\gamma$, the task is to identify two points $z_1$ and $z_2$ along $\mathbf{z}$ such that the integral in Equation~\ref{eq:p4} yields $\gamma$. Then $\theta_1 = z_1 \sigma  + \eta$ and $\theta_2 = z_2 \sigma + \eta$. In the literature, it is customary to assign the following values to the confidence coefficient: $\gamma = 0.95, 0.975, 0.99$.  

\begin{table*}[ht!]
    \centering
    \begin{tabular}{c|c|c|c|c|c|c|c|c|c}
    \hline
   Parameter & Distribution & $x_{\delta/2}$ & $x_{1-\delta/2}$ & CI (RR-Interval) & CI (HR) & Ref. (RR) & Ref. (HR) & p-value (RR) & p-value (HR)\\
    \hline     \hline
     $\eta$ & $N(0, 1)$ & $ -2.24$ & $2.24$ & $\left ( 0.73, 0.81 \right )$ & $\left ( 76.63, 84.79 \right )$ & $\sigma_{12} = 0.12$ &  $\sigma_{12} = 13.37$ & 0.23 & 0.10\\
    $\eta$ & $t(n-1)$ & $1.83$ & $2.63$ & $\left ( 0.76, 0.82 \right )$ & $\left ( 77.27, 85.66 \right )$ & $\mathbf{s} = 0.13$ & $\mathbf{s} = 13.83$ & 0.26 & 0.11 \\
    $\sigma^2$ & $\chi^2(54)$ & $35.59$ & $76.19$ & $\left ( 0.01, 0.02 \right )$ & $\left ( 126.70, 271.22 \right )$ &  $\mathbf{v} = 0.12$ & $\mathbf{v} = 13.37$ & - & - \\  
    $\sigma^2$ & $\chi^2(53)$ & $34.78$ & $75$  & $\left ( 0.01, 0.03 \right )$ & $\left ( 135.16, 291.47 \right )$ & $\mathbf{s} = 0.13$ & $\mathbf{s} = 13.83$ & - & -  \\
\hline
    \end{tabular}
    \caption{The $95\%$ confidence interval estimation for the means and variances of the RR-interval and HRV for the samples collected with the wireless electrocardiograms under the four different assumptions.}
    \label{tab:ci}
\end{table*}
  
\subsection{Unknown Mean with Known Variance}

In this subsection we determine the confidence interval for the mean of a features of interest based on the data collected with the wearable ECGs, taking the variance of the 12-lead ECGs, $\sigma_{12}^2$, as a reliable reference. Due to the  normality assumption, the sample mean in Equation~\ref{eq:p1} has the following parameters: $N(\eta, \sigma/\sqrt{n})$. Hence, the confidence interval for the sample mean can be expressed as follows:

\begin{equation}
    \label{eq:p5}
    P \left \{ \eta - z_1 \frac{\sigma_{12}}{\sqrt{n}} <  \widetilde{\mathbf{x}} < z_1 \frac{\sigma_{12}}{\sqrt{n} } \right \} =  G \left (  z_2 \right ) - G \left (  z_1 \right )
\end{equation}
Rearranging terms yields:

\begin{equation}
    \label{eq:p6}
    P \left \{ \widetilde{\mathbf{x}} - z_1 \frac{\sigma_{12}}{\sqrt{n}} <  \eta  < \widetilde{\mathbf{x}} + z_1 \frac{\sigma_{12}}{\sqrt{n} } \right \} =  G \left (  z_2 \right ) - G \left (  z_1 \right ) = \gamma
\end{equation}
Because of the symmetric nature of the standardized normal density at the origin, $z_1 = -z_2 = z$, so that $\int_{-\infty}^{-z} f(x) \; dx = \int_{z}^{\infty} f(x) \; dx = \delta/2$, where $\delta = 1 - \gamma$.

\subsection{Unknown Mean with Unknown Variance}

If we do not trust $\sigma_{12}^2$, the confidence we attach to the samples of the wearable ECG has to be determined from the samples themselves, in which case, we make use of Equation~\ref{eq:smean}. Still the assumption that $\widetilde{\mathbf{x}}$ is normal holds, but this time we take $\mathbf{s}^2$ to be a reliable estimate of the unknown variance. The estimated confidence interval is then:
\[
\left ( \widetilde{\mathbf{x}} + z_1 \frac{\mathbf{s}}{\sqrt{n}}, \widetilde{\mathbf{x}}  + z_2 \frac{\mathbf{s}}{\sqrt{n}},  \right )
\]
But now, we have two random variables to deal with and computing the confidence intervals should take their joint density function into account. In the literature, a third random variable is defined as a function of $\widetilde{\mathbf{x}}$ and $\mathbf{s}^2$ which has a Student $t$ distribution with $n-1$ degree of freedom \cite{ci1987confidence}:
\begin{equation}
    \label{eq:p7}
    \mathbf{t} = \frac{\widetilde{\mathbf{x}} - \eta}{\mathbf{s}/\sqrt{n}}
\end{equation}
This modifies our confidence interval definition as follows:
\begin{equation}
    \label{eq:p8}
   P \left \{ t_1 < \frac{\widetilde{\mathbf{x}} - \eta}{\mathbf{s}/\sqrt{n}} < t_2 \right \} = \gamma
\end{equation}
so that, the parameter we wish to estimate in terms of the wearable ECG samples becomes:

\begin{equation}
    \label{eq:p8}
   P \left \{ \widetilde{x } - t_1  s/\sqrt{n} <  \eta  <  \widetilde{x } + t_2   s/\sqrt{n} \right \} = \gamma
\end{equation}

\subsection{Unknown Variance with Known Mean}

As a third possibility, we take the mean of the 12-lead ECGs as a reliable reference and wish to estimate the confidence interval for the variance of a feature of interest based on the samples we collected using the wearable ECGs. Since we trust the mean of the 12-lead ECGs, we can replace Equation~\ref{eq:smean} with:
\begin{equation}
    \label{eq:var}
  \mathbf{v}^2 = \frac{1}{n} \sum_{i = 1}^n \left ( \mathbf{x}_i -  \eta_{12} \right )^2
\end{equation}
$E [ \widetilde{\mathbf{v}}^2  ] = \sigma^2$ and for a large $n$, its variance tends to zero. In the same way we standardize the normal distribution to simplify its integration, we can slightly modify Equation~\ref{eq:var} to simplify the determination of the confidence interval:
\begin{equation}
    \label{eq:var}
  \mathbf{y} = \frac{n  \mathbf{v}^2}{\sigma^2} = \sum_{i = 1}^n \left ( \frac{\mathbf{x}_i -  \eta_{12}}{\sigma} \right )^2
\end{equation}
$\mathbf{y} $ has the so-called Chi-square distribution, with $n$ degree of freedom, $\chi^2(n)$ \cite{papoulis2002probability}. Unlike the previous cases, this density is asymmetric and, therefore, we should determine the confidence interval by identifying two regions whose joint area, when integrated, yields the probability $\delta = 1- \gamma$:
\begin{align}
    \label{eq:p9}
    P \left \{ \frac{n \mathbf{v}^2}{\sigma^2} < z_1 \right \}  = \frac{\delta}{2}\\ \nonumber
    P \left \{ \frac{n \mathbf{v}^2}{\sigma^2} > z_2 \right \} = \frac{\delta}{2}
\end{align}
Hence, the confidence interval for the variance of a feature of interest is given us:
\begin{equation}
    \label{eq:p10}
  \left (  \frac{n v^2}{\chi_{1-\delta/2}(n)} < \sigma^2 < \frac{n v^2}{\chi_{\delta/2}(n)} \right ) = \gamma
\end{equation}
where $\chi_{q}(n)$ is the chi-square's $q$ percentile of the random variable $\mathbf{y}$. 

\subsection{Unknown Variance with Unknown Mean}
Lastly, we attempt to estimate the uncertainty associated with the samples of the wearable ECGs without taking the statistics of the 12-lead ECGs as a reference altogether. In this case the sample variance (Equation~\ref{eq:smean}) is used as a reference. Similar to the density of $\mathbf{v}^2$, normalizing $\mathbf{s}^2$ simplifies the computation of the confidence interval:
\begin{equation}
    \label{eq:var}
  \mathbf{w} =   \frac{(n - 1) \mathbf{s}^2}{\sigma^2} = \sum_{i = 1}^n \left ( \frac{\mathbf{x}_i -  \widetilde{\mathbf{x}}}{\sigma} \right )^2
\end{equation}
$\mathbf{w}$ has a chi-square distribution of degree $n-1$: $\chi^2 (n - 1)$. With this, the  confidence interval for the unknown variance is given as:
\begin{equation}
    \label{eq:11}
    P \left \{ \chi_{\delta/2}^2(n-1) < \frac{(n-1) \mathbf{s}^2}{\sigma^2} < \chi^2 _{1 - \delta/2}(n-1)  \right \} = \gamma
\end{equation}
This yields the following interval:
\begin{equation}
    \label{eq:12}
    P \left \{ \frac{(n-1)s^2}{\chi_{1-\delta/2}^2(n-1)} < \sigma^2  < \frac{(n-1)s^2}{\chi^2 _{\delta/2}(n-1)}  \right \} = \gamma
\end{equation}

\subsection{Interval Estimates}

Table \ref{tab:parameter} shows the 95\% confidence intervals for the RR interval and HRV based on the four different assumptions regarding the reference values for the interval estimates. The calculated p-values suggest that the hypothesis $\mathbf{H}_0$, that the statistics obtained using wireless ECGs significantly overlap with those obtained using 12-lead ECGs, is not sufficiently supported to be rejected. With 95\% confidence, the mean RR interval lies between 0.73 and 0.81, assuming the variance of the 12-lead ECGs can be used as a reliable reference. In this case, the p-value for the hypothesis that the mean RR interval obtained using the 12-lead statistics and the wireless ECG statistics is identical is 0.23. If we reject the variance of the 12-lead ECG as a reliable reference, the mean RR interval lies between 0.76 and 0.82. The corresponding p-value is 0.26. In both cases, the hypothesis is confirmed. The same applies to HRV. We obtained a p-value of 0.10 assuming that the validity determined using the 12-lead ECG statistics is reliable; and a p-value of 0.11 when this assumption is not true.

\section{Activity vs Heart Rate Variation}
\label{sec:hrv}
Heart rate variability is an important parameter for cardiac and autonomic fitness. Although it can be influenced by various physiological factors and age, it is nevertheless a reliable indicator of underlying autonomic function and has independent prognostic values for the early detection of various heart diseases in the general population \cite{natarajan2020heart}.  
\begin{figure*}
   \centering
\includegraphics[width=0.8\textwidth]{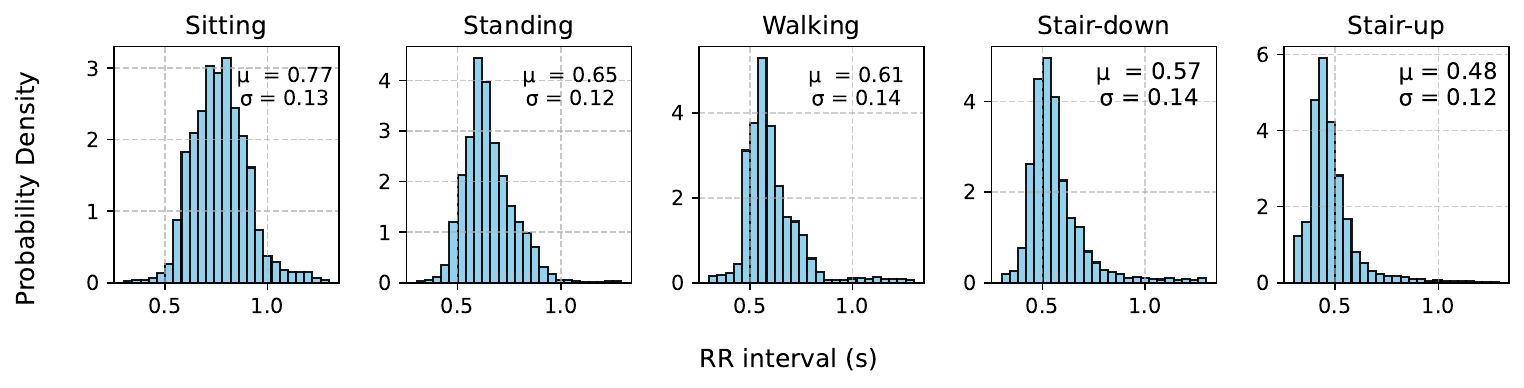}
    \centering
\includegraphics[width=0.8\textwidth]{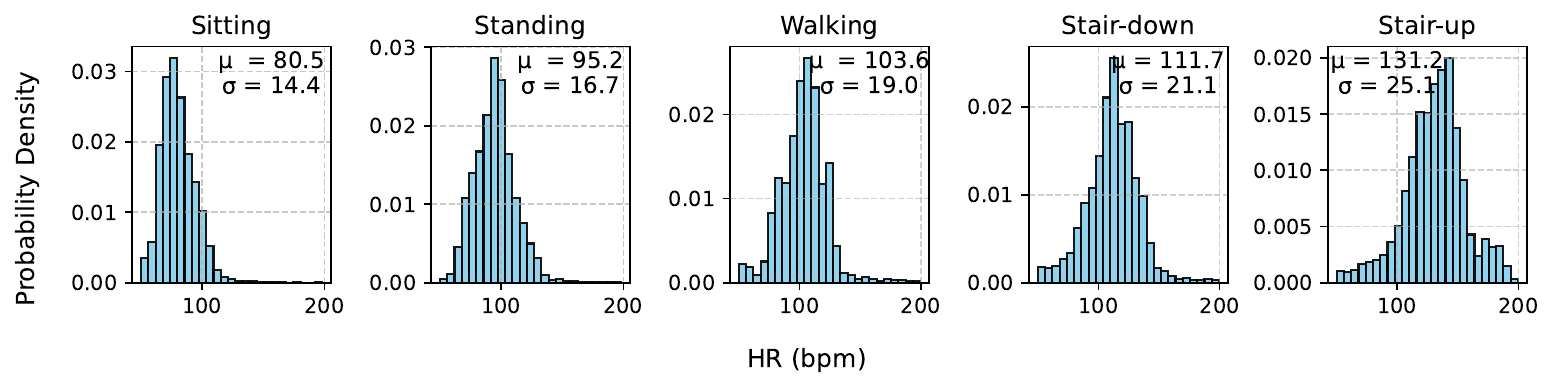}
\caption{RR-interval and HRV statistics for different activities.}
\label{fig:rr-hr-activities}
\end{figure*}
Fig.~\ref{fig:rr-hr-activities} displays the RR-interval and heart rate statistics of the five different physical activities (sitting, standing, walking, climbing down stairs, and climbing up stairs). The statistics represent the Shimmer wireless electrocardiogram dataset involving  54 subjects carrying out different physical activities. A closer inspection of the means and standard deviations of the distributions reveals that RR interval decreases and heart rate increases with increase in the intensity of the underlying physical activities. 
In the literature, there are different approaches to analyze long-term and short-term heart rate variability, which includes time-domain analysis, non-linear analysis, and frequency-domain analysis. In the following subsections, we apply these techniques to closely investigate the dynamics between physical exertion and heart rate variability. 
\subsubsection{Time-domain analysis}
The RR-interval sequence reveals instantaneous heart rate variability. In some situations, this revelation alone may not be sufficient to reason about what causes it. Among the common metrics used to establish quasi-stable temporal aspects are the standard deviation of all normal-to-normal intervals (SDNN) and the root mean square of successive differences (RMSSD), which reflect overall variability and short-term changes, respectively \cite{shaffer2017overview}.  In clinical settings, low SDNN (particularly, in long-term ECG recordings) is linked to abnormal cardiac conditions, and lower RMSSD indicates reduced vagal activity (particularly, due to stress) \cite{shaffer2017overview}.
\begin{align}
\label{eq:hrv1}
SDNN &= \sqrt{\frac{1}{N-1} \sum_{i=1}^{N} (RR_i - \overline{RR})^2} \\
RMSSD &= \sqrt{\frac{1}{N-1} \sum_{i=1}^{N-1} (RR_{i+1} - RR_i)^2}
\end{align}
where: $RR_i$ refers to the $i$-th RR interval; $\overline{RR}$, to the mean of all RR intervals; $N$, to the total number of RR intervals; and $RR_{i+1} - RR_i$, to the difference between successive RR intervals. 
For our subjects, the RR interval  and HR vary across the different activities;  RR interval is long during sitting and relatively short during intense physical activity, indicating the expected increase in heart rate  to meet the body’s increased  energy demands. As shown in Figs.~\ref{fig:22},  during sitting, the histogram of RR interval is wider, suggesting the RR interval varies more significantly, resulting a high SDNN and a high RMSSD. As the activity's exertion level increases, the RR interval histogram progressively becomes narrower, suggesting low overall variability, which corresponds to low SDNN and low RMSSD values. From this it can be confirmed that analyzing time domain parameters for the different physical activities enables the early detection of activity dependent cardiac abnormalities and understanding how physical activity influences cardiovascular function.
\subsubsection{Nonlinear analysis}
\begin{figure}
    \centering
\includegraphics[width=0.45\textwidth]{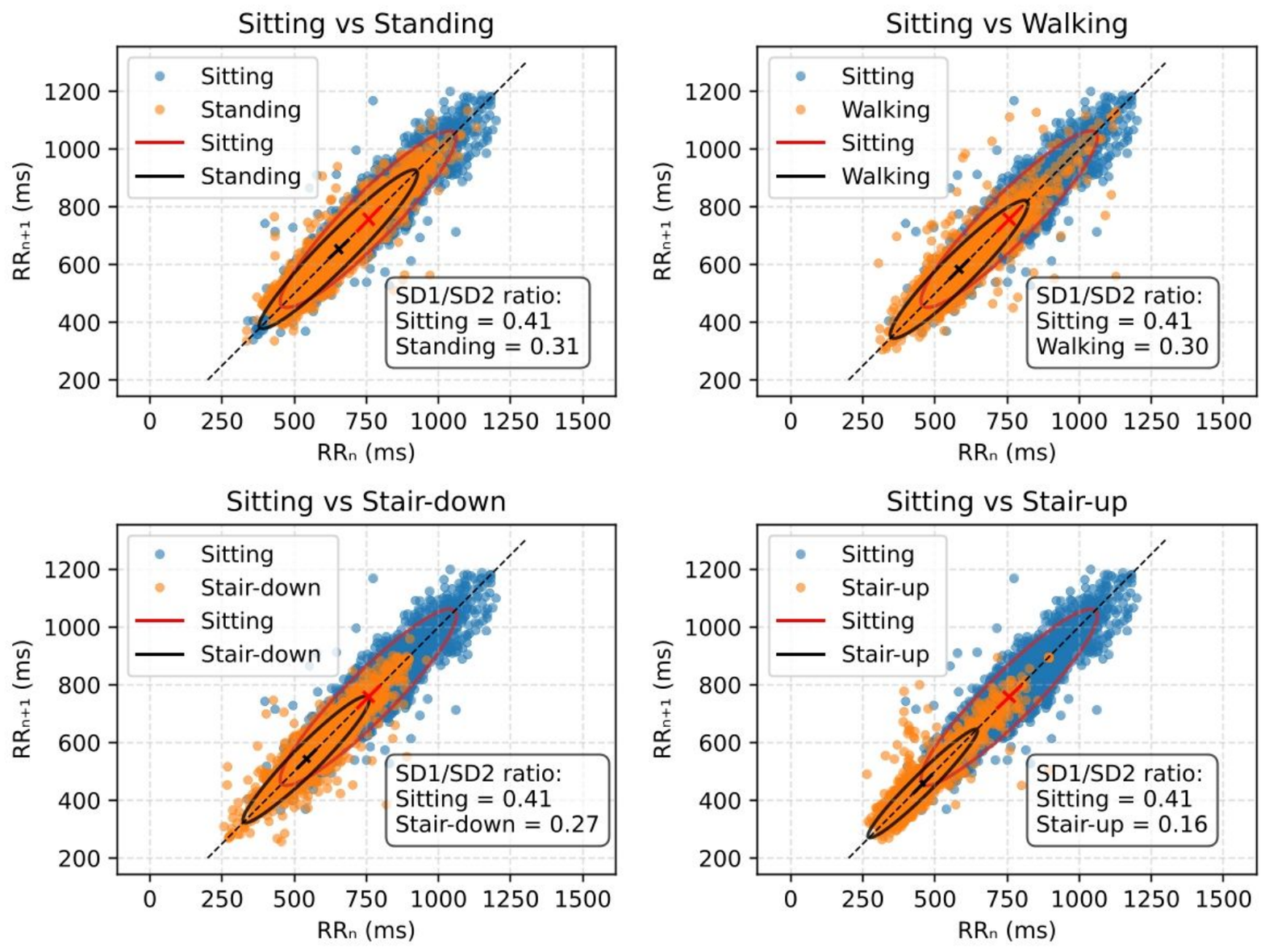}
\caption{Poincaré plots, comparing the cardiac features of sitting with the other activities.}
\label{fig:22p}
\end{figure}

\begin{figure}
    \centering
\includegraphics[width=0.45\textwidth]{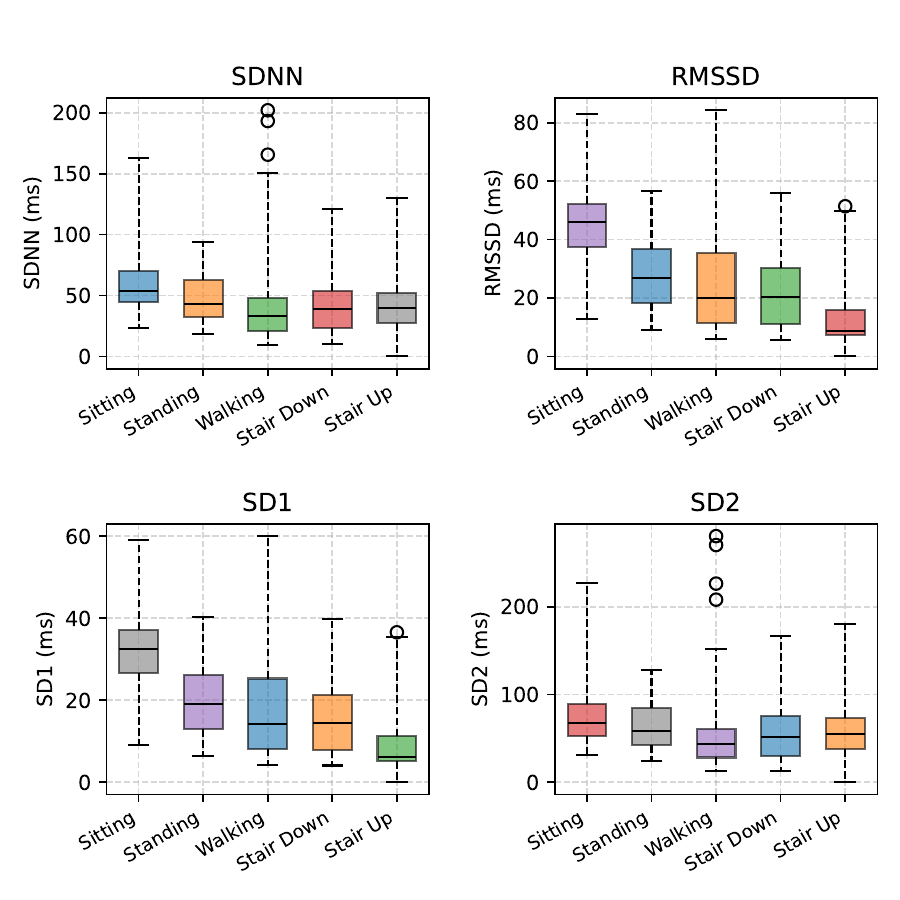}
\caption{Boxplot of HRV metrics (SDNN, RMSSD, SD1, SD2) for different activities.}
\label{fig:22}
\end{figure}

Non-linear analysis is employed to determine more complex and nonlinear heart rate variability (due to stress, complex physical exercise, and some cardiac diseases) which are otherwise challenging to diagnose. These include Poincaré plots, approximate entropy , and detrended fluctuation \cite{nardelli2015recognizing}. In this study Poincaré plots, a method providing a plot of Short-term variability (SD1) and long-term variability (SD2), adopted. Unlike the approaches we adopted so far, for this study we consider the statistics of individual subjects, which we randomly selected from our 54 subjects. For these individual, the RR-interval, short-term variability (SD1) and long-term variability (SD2) are computed. Afterwards, the mean of these parameters are used to produce the Poincaré plots. 
Fig.~\ref{fig:22p} shows that during a resting state (sitting), the Poincaré plot is wider, highlighting a high HRV and a strong vagal (parasympathetic) activity. During a physical exertion, the ellipse  becomes narrower and more elongated along the line of identity, indicating a reduction in the short-term variability and a shift toward sympathetic predominance. When the physical exertion become more intense (climbing up a stair), the ellipse becomes very thin and aligned with the line of identity, reflecting a low overall variability and dominant sympathetic control. 
The SD1/SD2 ratio in a Poincaré plot represents the relationships between short-term heart rate variability (SD1) and long-term heart rate variability (SD2); a higher ratio suggesting a dominant parasympathetic influence, while a lower ratio suggesting a sympathetic predominance. As shown in figure\ref{fig:22p}, while sitting, the ratio is high (0.41); during a moderate exertion (such as standing, walking, and climbing down a staircase), the SD1/SD2 ratio begins to decrease (becoming  0.31, 0.30 and 0.27, respectively). During a more intense physical exertion (climbing up a staircase), the ratio reaches its lowest values of 0.16.
\begin{align}
\label{eq:hrv2}
SD1 &= \sqrt{\frac{1}{2} Var(RR_{i+1} - RR_i)} \\
SD2 &= \sqrt{2 SDNN^2 -  SD1^2}
\end{align}
Alternatively, the long-term (SDNN and SD2) and the short-term (RMSSD and SD1) HRV parameters can be compared with box plots. Figure \ref{fig:22} shows that the median of all HRV parameters is highest when sitting and becomes the lowest during climbing up stairs, reflecting the progressive withdrawal of parasympathetic dominance and increasing sympathetic influence.  The width of the boxes is wider during walking indicating greater inter-personal variability (i.e., the participants cardiovascular system responded differently to walking), whereas the width of the boxes  of all HRV parameters is the narrowest during climbing up stairs, suggesting a more consistent autonomic response across the subjects.
Table \ref{tab:hrv_metrics} summarize the means and standard deviations  of both long-term and short-term HRV parameters across different activities.  At resting condition, such as sitting and standing, the mean values of both long-term and short-term parameters are high, indicating greater overall heart rate variability.  As activity intensity increase, like walking, climbing down and climbing up stairs, the mean values of both long-term and short-term parameters decrease,  especially short-term parameters, such as RMSSD and SD1, tend to decrease more rapidly since they are primarily driven by parasympathetic activity.  However, long-term parameters like SDNN and SD2  decrease moderately, as they reflect the alternation between parasympathetic and sympathetic activity.
Overall, the standard deviations of the different parameters does not vary appreciably across different activities. It is, nevertheless, relatively small for the short-term parameters ( RMSSD and SD1). In other words, short-term HRV is relatively stable among individuals, regardless of the underlying physical exertion. By contrast, the standard deviations of the long-term  parameters (SDNN and SD2) are relatively high across all activities, suggesting an appreciable individual difference in HRV. Compared to the other physical activities, the standard deviations of the short- and long-term  parameters are higher for walking. This can be due to differences in walking intensity and pace. This said, it is worth mentioning that all these parameters can be affected by various factors including age, physical fitness, and device sensitivity. 
\begin{table}[h!]
\centering
\begin{tabular}{lcccc}
\hline
\textbf{Activity} & \textbf{SDNN (ms)} & \textbf{RMSSD (ms)} & \textbf{SD1 (ms)} & \textbf{SD2 (ms)} \\
\hline
Sitting     & 59.4 $\pm$ 24.8 & 44.6$\pm$ 13.9 & 31.6 $\pm$ 9.9& 77.1$\pm$ 35.2 \\
Standing    & 46.8 $\pm$ 19.4 & 27.9 $\pm$ 12.2 & 19.8 $\pm$ 8.7 & 62.9 $\pm$ 26.8 \\
Walking     & 45.8 $\pm$ 45.7 & 25.8 $\pm$ 18.9 & 18.3 $\pm$ 13.4  & 61.7 $\pm$ 63.7 \\
Stair Down  & 43.6 $\pm$ 26.9 & 22.6 $\pm$ 13.3 & 16.0$\pm$ 9.4 & 59.0 $\pm$ 37.7 \\
Stair Up    & 43.0 $\pm$ 26.2 & 13.7$\pm$ 11.2& 9.7 $\pm$ 7.9 & 59.9 $\pm$ 36.3\\
\hline
\end{tabular}
\caption{HRV metrics for different activities (mean $\pm$ SD).}
\label{tab:hrv_metrics}
\end{table}
\subsubsection{Frequency-domain analysis}
\begin{figure}[H]
    \centering
\includegraphics[width=0.45\textwidth]{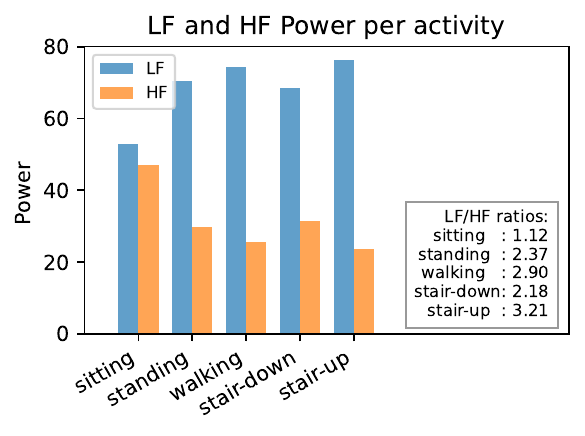}
\caption{LF and HF power across different activity.}
\label{fig:lfhfnew}
\end{figure}

The spectral analysis of the RR-interval is often used to estimate the effect of sympathetic and parasympathetic modulation of the RR-interval \cite{mcsharry2003dynamical}. Unlike many electrical signals, however, the RR-interval, which is intrinsically sampled at irregular interval, does not represent a uniformly sampled sequence. Therefore, the usual Fourier Transform is not an ideal transformation to reveal the spectral aspect of the HRV. Instead, we apply the Lomb–Scargle periodogram \cite{fonseca2013lomb}, which involves standardization, detrending, and the application of Welch method \cite{barbe2009welch} on overlapping segments of the RR-interval. The two dominant frequency bands associated with the sympathetic and parasympathetic activity are referred to as the low-frequency (LF) band (ca. 0.04–0.15 Hz) and the high-frequency (HF) band (ca. 0.15–0.4 Hz), respectively, \cite{mcsharry2003dynamical}. 
\begin{figure}[H]
    \centering
\includegraphics[width=0.45\textwidth]{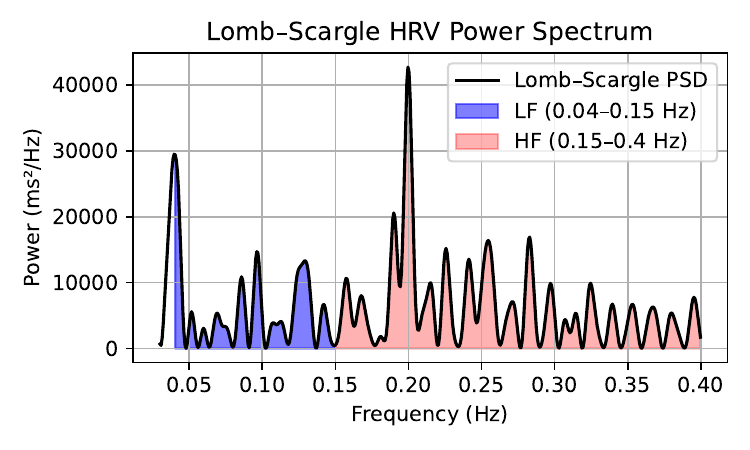}
\includegraphics[width=0.45\textwidth]{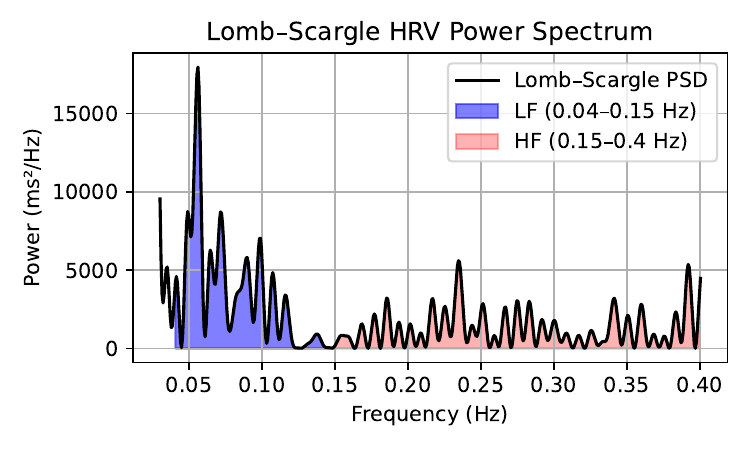}
\caption{The power spectrum of LF and HF components for sitting (above) climbing up staircase (bottom).}
\label{fig:hflf}
\end{figure}
In response to physical exertion or mental stress, sympathetic activity (LF power) increases, while parasympathetic activity (HF power) decreases \cite{hernando2016inclusion}. Our statistics seems to confirm this trend as the power spectra of a randomly selected individual exhibits in Figs.~\ref{fig:lfhfnew} and ~\ref{fig:hflf}. Accordingly, in a resting state (sitting), a higher HF power is observed, whereas the LF power becomes dominant during physical exertion (climbing up a staircase) proportionally increasing  as HF power diminishes. 
HRV spectral power expressed in absolute units can be  distorted by  overall spectral power changes,  and  LF and HF power expressed in normalized (relative) units provides a more realistic reflection of autonomic balance \cite{mcsharry2003dynamical}. Figure \ref{fig:lfhfnew} present the  normalized average values of LF and HF for all the subjects in the Shimmer dataset; normalization is done per subject and then the average values of the normalized LF and HF are computed. The LF spectral power was relatively low during sitting,  and it increased progressively as the intensity of the physical exertion increases,  whereas the HF spectral power was relatively high during sitting and decreases as the activity level increase. Correspondingly,the LF/HF ratio was low during sitting, suggesting greater parasympathetic predominance under resting conditions, and the LF/HF ratio was high during walking and  climbing up stairs, illustrating increased sympathetic dominance.

\section{Conclusions}
\label{sec:conclusion}

In this study, we investigated the quality of data collected by wireless electrocardiograms outside clinical environments, without medical supervision. We used data collected by medically certified 12-lead ECGs and Holter ambulatory ECGs, collected in clinical settings or under medical supervision, for comparison. Our focus was on the RR-interval time series and heart rate variability, as the essential ECG features for these features can be identified with relative ease in all the datasets. Once the histograms of the RR-intervals and heart rate variability were established, we compared their overlaps and computed confidence intervals based on four different assumptions. The statistics suggest that there is a considerable overlap between the different datasets. In addition, we investigated the relationship between physical exertion and cardiac response, subjecting our participants to physical activities with variable degree of intensity. The various analytic approaches we adopted to the investigation suggest that a relatively high heart rate variability can be observed in a resting (sitting) condition, whereas as the intensity level of the underlying physical exertion increases, the heart rate variability decreases considerably. 

This study provides valuable insights into the dynamics between sympathetic and parasympathetic activity (the so-called sympathovagal balance) and into threshold values for differentiating various heart conditions/workloads. In the future, we plan to expand our dataset and employ different types of wireless electrocardiograms. Furthermore, we aim to differentiate the various potential sources of error (e.g., intrinsic device inaccuracies, calibration error, peak detection errors, placement errors).


\begin{thebibliography}{10}
\providecommand{\url}[1]{#1}
\csname url@samestyle\endcsname
\providecommand{\newblock}{\relax}
\providecommand{\bibinfo}[2]{#2}
\providecommand{\BIBentrySTDinterwordspacing}{\spaceskip=0pt\relax}
\providecommand{\BIBentryALTinterwordstretchfactor}{4}
\providecommand{\BIBentryALTinterwordspacing}{\spaceskip=\fontdimen2\font plus
\BIBentryALTinterwordstretchfactor\fontdimen3\font minus
  \fontdimen4\font\relax}
\providecommand{\BIBforeignlanguage}[2]{{%
\expandafter\ifx\csname l@#1\endcsname\relax
\typeout{** WARNING: IEEEtran.bst: No hyphenation pattern has been}%
\typeout{** loaded for the language `#1'. Using the pattern for}%
\typeout{** the default language instead.}%
\else
\language=\csname l@#1\endcsname
\fi
#2}}
\providecommand{\BIBdecl}{\relax}
\BIBdecl

\bibitem{masihi2021development}
S.~Masihi, M.~Panahi, D.~Maddipatla, A.~J. Hanson, S.~Fenech, L.~Bonek,
  N.~Sapoznik, P.~D. Fleming, B.~J. Bazuin, and M.~Z. Atashbar, ``Development
  of a flexible wireless ecg monitoring device with dry fabric electrodes for
  wearable applications,'' \emph{IEEE Sensors Journal}, vol.~22, no.~12, pp.
  11\,223--11\,232, 2021.

\bibitem{chen2022contactless}
J.~Chen, D.~Zhang, Z.~Wu, F.~Zhou, Q.~Sun, and Y.~Chen, ``Contactless
  electrocardiogram monitoring with millimeter wave radar,'' \emph{IEEE
  Transactions on Mobile Computing}, vol.~23, no.~1, pp. 270--285, 2022.

\bibitem{farooq2021wearable}
A.~Farooq, M.~Seyedmahmoudian, and A.~Stojcevski, ``A wearable wireless sensor
  system using machine learning classification to detect arrhythmia,''
  \emph{IEEE Sensors Journal}, vol.~21, no.~9, pp. 11\,109--11\,116, 2021.

\bibitem{farrokhi2024human}
S.~Farrokhi, W.~Dargie, and C.~Poellabauer, ``Human activity recognition based
  on wireless electrocardiogram and inertial sensors,'' \emph{IEEE Sensors
  Journal}, vol.~24, no.~5, pp. 6490--6499, 2024.

\bibitem{strodthoff2020deep}
N.~Strodthoff, P.~Wagner, T.~Schaeffter, and W.~Samek, ``Deep learning for ecg
  analysis: Benchmarks and insights from ptb-xl,'' \emph{IEEE journal of
  biomedical and health informatics}, vol.~25, no.~5, pp. 1519--1528, 2020.

\bibitem{gomez2021platform}
C.~A. G{\'o}mez-Garc{\'\i}a, M.~Askar-Rodriguez, and J.~Velasco-Medina,
  ``Platform for healthcare promotion and cardiovascular disease prevention,''
  \emph{IEEE Journal of Biomedical and Health Informatics}, vol.~25, no.~7, pp.
  2758--2767, 2021.

\bibitem{hussain2021big}
I.~Hussain and S.~J. Park, ``Big-ecg: Cardiographic predictive cyber-physical
  system for stroke management,'' \emph{IEEe Access}, vol.~9, pp.
  123\,146--123\,164, 2021.

\bibitem{breen2022ecg}
C.~Breen, G.~Kelly, and W.~Kernohan, ``Ecg interpretation skill acquisition: A
  review of learning, teaching and assessment,'' \emph{Journal of
  electrocardiology}, vol.~73, pp. 125--128, 2022.

\bibitem{houghton2025making}
A.~Houghton, \emph{Making sense of the ECG: a hands-on guide}.\hskip 1em plus
  0.5em minus 0.4em\relax CRC press, 2025.

\bibitem{bayoumy2021smart}
K.~Bayoumy, M.~Gaber, A.~Elshafeey, O.~Mhaimeed, E.~H. Dineen, F.~A. Marvel,
  S.~S. Martin, E.~D. Muse, M.~P. Turakhia, K.~G. Tarakji \emph{et~al.},
  ``Smart wearable devices in cardiovascular care: where we are and how to move
  forward,'' \emph{Nature Reviews Cardiology}, vol.~18, no.~8, pp. 581--599,
  2021.

\bibitem{fujiwara2018heart}
K.~Fujiwara, E.~Abe, K.~Kamata, C.~Nakayama, Y.~Suzuki, T.~Yamakawa,
  T.~Hiraoka, M.~Kano, Y.~Sumi, F.~Masuda \emph{et~al.}, ``Heart rate
  variability-based driver drowsiness detection and its validation with eeg,''
  \emph{IEEE transactions on biomedical engineering}, vol.~66, no.~6, pp.
  1769--1778, 2018.

\bibitem{dogan2025continuous}
A.~Dogan, A.~Bishnoi, R.~B. Sowers, and M.~E. Hernandez, ``Continuous heart
  rate recovery monitoring with ecg signals from wearables: Identifying risk
  groups in the general population,'' \emph{IEEE Journal of Biomedical and
  Health Informatics}, 2025.

\bibitem{elghozi2007sympathetic}
J.-L. Elghozi and C.~Julien, ``Sympathetic control of short-term heart rate
  variability and its pharmacological modulation,'' \emph{Fundamental \&
  clinical pharmacology}, vol.~21, no.~4, pp. 337--347, 2007.

\bibitem{rohleder2004psychosocial}
N.~Rohleder, U.~M. Nater, J.~M. Wolf, U.~Ehlert, and C.~Kirschbaum,
  ``Psychosocial stress-induced activation of salivary alpha-amylase: an
  indicator of sympathetic activity?'' \emph{Annals of the New York Academy of
  Sciences}, vol. 1032, no.~1, pp. 258--263, 2004.

\bibitem{fleshner2005physical}
F.~Fleshner, ``Physical activity and stress resistance: sympathetic nervous
  system adaptations prevent stress-induced immunosuppression,'' \emph{Exercise
  and sport sciences reviews}, vol.~33, no.~3, pp. 120--126, 2005.

\bibitem{strigo2016interoception}
I.~A. Strigo and A.~D. Craig, ``Interoception, homeostatic emotions and
  sympathovagal balance,'' \emph{Philosophical Transactions of the Royal
  Society B: Biological Sciences}, vol. 371, no. 1708, p. 20160010, 2016.

\bibitem{di2016does}
D.~Di~Raimondo, G.~Miceli, A.~Casuccio, A.~Tuttolomondo, C.~Butta, V.~Zappulla,
  C.~Schimmenti, G.~Musiari, and A.~Pinto, ``Does sympathetic overactivation
  feature all hypertensives? differences of sympathovagal balance according to
  night/day blood pressure ratio in patients with essential hypertension,''
  \emph{Hypertension Research}, vol.~39, no.~6, pp. 440--448, 2016.

\bibitem{fuller2020reliability}
D.~Fuller, E.~Colwell, J.~Low, K.~Orychock, M.~A. Tobin, B.~Simango, R.~Buote,
  D.~Van~Heerden, H.~Luan, K.~Cullen \emph{et~al.}, ``Reliability and validity
  of commercially available wearable devices for measuring steps, energy
  expenditure, and heart rate: systematic review,'' \emph{JMIR mHealth and
  uHealth}, vol.~8, no.~9, p. e18694, 2020.

\bibitem{zang2025novel}
J.~Zang, Q.~An, B.~Li, Z.~Zhang, L.~Gao, and C.~Xue, ``A novel wearable device
  integrating ecg and pcg for cardiac health monitoring,'' \emph{Microsystems
  \& Nanoengineering}, vol.~11, no.~1, p.~7, 2025.

\bibitem{rafols2018evaluation}
M.~Rafols-de Urquia, L.~Estrada, J.~Estevez-Piorno, L.~Sarlabous, R.~Jane, and
  A.~Torres, ``Evaluation of a wearable device to determine cardiorespiratory
  parameters from surface diaphragm electromyography,'' \emph{IEEE journal of
  biomedical and health informatics}, vol.~23, no.~5, pp. 1964--1971, 2018.

\bibitem{lazaro2020wearable}
J.~L{\'a}zaro, N.~Reljin, M.-B. Hossain, Y.~Noh, P.~Laguna, and K.~H. Chon,
  ``Wearable armband device for daily life electrocardiogram monitoring,''
  \emph{IEEE Transactions on Biomedical Engineering}, vol.~67, no.~12, pp.
  3464--3473, 2020.

\bibitem{bent2020investigating}
B.~Bent, B.~A. Goldstein, W.~A. Kibbe, and J.~P. Dunn, ``Investigating sources
  of inaccuracy in wearable optical heart rate sensors,'' \emph{NPJ digital
  medicine}, vol.~3, no.~1, p.~18, 2020.

\bibitem{natarajan2020heart}
A.~Natarajan, A.~Pantelopoulos, H.~Emir-Farinas, and P.~Natarajan, ``Heart rate
  variability with photoplethysmography in 8 million individuals: a
  cross-sectional study,'' \emph{The Lancet Digital Health}, vol.~2, no.~12,
  pp. e650--e657, 2020.

\bibitem{betti2017evaluation}
S.~Betti, R.~M. Lova, E.~Rovini, G.~Acerbi, L.~Santarelli, M.~Cabiati,
  S.~Del~Ry, and F.~Cavallo, ``Evaluation of an integrated system of wearable
  physiological sensors for stress monitoring in working environments by using
  biological markers,'' \emph{IEEE transactions on biomedical engineering},
  vol.~65, no.~8, pp. 1748--1758, 2017.

\bibitem{8638779}
R.~Bhardwaj and V.~Balasubramanian, ``Viability of cardiac parameters measured
  unobtrusively using capacitive coupled electrocardiography (cecg) to estimate
  driver performance,'' \emph{IEEE Sensors Journal}, vol.~19, no.~11, pp.
  4321--4330, 2019.

\bibitem{li2025motion}
D.~Li, T.-R. Cui, J.-H. Liu, W.-C. Shao, X.~Liu, Z.-K. Chen, Z.-G. Xu, X.~Li,
  S.-Y. Xu, Z.-Y. Xie \emph{et~al.}, ``Motion-unrestricted dynamic
  electrocardiogram system utilizing imperceptible electronics,'' \emph{Nature
  Communications}, vol.~16, no.~1, p. 3259, 2025.

\bibitem{burns2010shimmer}
A.~Burns, B.~R. Greene, M.~J. McGrath, T.~J. O'Shea, B.~Kuris, S.~M. Ayer,
  F.~Stroiescu, and V.~Cionca, ``Shimmer™--a wireless sensor platform for
  noninvasive biomedical research,'' \emph{IEEE Sensors Journal}, vol.~10,
  no.~9, pp. 1527--1534, 2010.

\bibitem{wagner2020ptb}
P.~Wagner, N.~Strodthoff, R.-D. Bousseljot, D.~Kreiseler, F.~I. Lunze,
  W.~Samek, and T.~Schaeffter, ``Ptb-xl, a large publicly available
  electrocardiography dataset,'' \emph{Scientific data}, vol.~7, no.~1, pp.
  1--15, 2020.

\bibitem{makowski2021neurokit2}
D.~Makowski, T.~Pham, Z.~J. Lau, J.~C. Brammer, F.~Lespinasse, H.~Pham,
  C.~Sch{\"o}lzel, and S.~A. Chen, ``Neurokit2: A python toolbox for
  neurophysiological signal processing,'' \emph{Behavior research methods},
  vol.~53, no.~4, pp. 1689--1696, 2021.

\bibitem{martinez2004wavelet}
J.~P. Mart{\'\i}nez, R.~Almeida, S.~Olmos, A.~P. Rocha, and P.~Laguna, ``A
  wavelet-based ecg delineator: evaluation on standard databases,'' \emph{IEEE
  Transactions on biomedical engineering}, vol.~51, no.~4, pp. 570--581, 2004.

\bibitem{sangha2022automated}
V.~Sangha, B.~J. Mortazavi, A.~D. Haimovich, A.~H. Ribeiro, C.~A. Brandt, D.~L.
  Jacoby, W.~L. Schulz, H.~M. Krumholz, A.~L.~P. Ribeiro, and R.~Khera,
  ``Automated multilabel diagnosis on electrocardiographic images and
  signals,'' \emph{Nature communications}, vol.~13, no.~1, p. 1583, 2022.

\bibitem{rahul2021improved}
J.~Rahul, M.~Sora, L.~D. Sharma, and V.~K. Bohat, ``An improved cardiac
  arrhythmia classification using an rr interval-based approach,''
  \emph{Biocybernetics and Biomedical Engineering}, vol.~41, no.~2, pp.
  656--666, 2021.

\bibitem{mcsharry2003dynamical}
P.~E. McSharry, G.~D. Clifford, L.~Tarassenko, and L.~A. Smith, ``A dynamical
  model for generating synthetic electrocardiogram signals,'' \emph{IEEE
  transactions on biomedical engineering}, vol.~50, no.~3, pp. 289--294, 2003.

\bibitem{yang2021mean}
J.~Yang, S.~Rahardja, and P.~Fr{\"a}nti, ``Mean-shift outlier detection and
  filtering,'' \emph{Pattern Recognition}, vol. 115, p. 107874, 2021.

\bibitem{blazquez2021review}
A.~Bl{\'a}zquez-Garc{\'\i}a, A.~Conde, U.~Mori, and J.~A. Lozano, ``A review on
  outlier/anomaly detection in time series data,'' \emph{ACM computing surveys
  (CSUR)}, vol.~54, no.~3, pp. 1--33, 2021.

\bibitem{Dogan10921692}
A.~Dogan, A.~Bishnoi, R.~B. Sowers, and M.~E. Hernandez, ``Continuous heart
  rate recovery monitoring with ecg signals from wearables: Identifying risk
  groups in the general population,'' \emph{IEEE Journal of Biomedical and
  Health Informatics}, vol.~29, no.~8, pp. 5493--5502, 2025.

\bibitem{monkaresi2013machine}
H.~Monkaresi, R.~A. Calvo, and H.~Yan, ``A machine learning approach to improve
  contactless heart rate monitoring using a webcam,'' \emph{IEEE journal of
  biomedical and health informatics}, vol.~18, no.~4, pp. 1153--1160, 2013.

\bibitem{papoulis2002probability}
A.~Papoulis and S.~U. Pillai, \emph{Probability}.\hskip 1em plus 0.5em minus
  0.4em\relax McGraw-hill, 2002.

\bibitem{ci1987confidence}
B.~Ci and R.-O. Rule, ``Confidence intervals,'' \emph{Lancet}, vol.~1, no.
  8531, pp. 494--7, 1987.

\bibitem{shaffer2017overview}
F.~Shaffer and J.~P. Ginsberg, ``An overview of heart rate variability metrics
  and norms,'' \emph{Frontiers in public health}, vol.~5, p. 258, 2017.

\bibitem{nardelli2015recognizing}
M.~Nardelli, G.~Valenza, A.~Greco, A.~Lanata, and E.~P. Scilingo, ``Recognizing
  emotions induced by affective sounds through heart rate variability,''
  \emph{IEEE Transactions on Affective Computing}, vol.~6, no.~4, pp. 385--394,
  2015.

\bibitem{fonseca2013lomb}
D.~S. Fonseca, A.~Netto, R.~B. Ferreira, and A.~M. De~Sa, ``Lomb-scargle
  periodogram applied to heart rate variability study,'' in \emph{2013 ISSNIP
  Biosignals and Biorobotics Conference: Biosignals and Robotics for Better and
  Safer Living (BRC)}.\hskip 1em plus 0.5em minus 0.4em\relax IEEE, 2013, pp.
  1--4.

\bibitem{barbe2009welch}
K.~Barbe, R.~Pintelon, and J.~Schoukens, ``Welch method revisited:
  nonparametric power spectrum estimation via circular overlap,'' \emph{IEEE
  Transactions on signal processing}, vol.~58, no.~2, pp. 553--565, 2009.

\bibitem{hernando2016inclusion}
A.~Hernando, J.~Lazaro, E.~Gil, A.~Arza, J.~M. Garz{\'o}n, R.~Lopez-Anton,
  C.~De~La~Camara, P.~Laguna, J.~Aguil{\'o}, and R.~Bail{\'o}n, ``Inclusion of
  respiratory frequency information in heart rate variability analysis for
  stress assessment,'' \emph{IEEE journal of biomedical and health
  informatics}, vol.~20, no.~4, pp. 1016--1025, 2016.

\end{thebibliography}
\end{document}